\documentstyle[12pt,qqa4lart]{article}
\input tcilatex

\QQQ{Language}{
American English
}

\begin{document}

\author{Lu-Ming Duan and Guang-Can Guo \\
%EndAName
Department of Physics and Nonlinear Science Center,\\
University of Science and Technology of China,\\
Hefei 230026, People's Republic of China}
\title{Reply to the Comment }
\date{}
\maketitle

\baselineskip 18pt

{\bf PACS numbers:} 89.70.+c, 03.65.Bz, 42.50.Dv\\

In the comment [1], Zanardi and Rasetti argue that several claims in our
recent letter [2] are questionable. Here we show these claims remain true.
Our points are that (i) the Hamiltonian (1) in Ref. [2] describes amplitude
damping as well as phase damping; (ii) the free-Hamiltonian-elimination
(FHE) has feasible implementation in practice and there exists a simple FHE
procedure which is designed without the knowledge of the noise parameters $%
\lambda ^{\left( i\right) }$; (iii) the gate operation constructed in the
letter [2] is an encoded universal operation, not an encoded controlled NOT
(CNOT).

(i) There are mainly two kinds of dissipation, amplitude damping and phase
damping. General amplitude damping of many qubits is usually described by
the following interaction Hamiltonian [3] 
\begin{equation}
\label{1}H_I=\stackunder{l}{\sum }\left( g_l^{*}\sigma _l^{+}B_l+g_l\sigma
_l^{-}B_l^{+}\right) , 
\end{equation}
where all $B_l$ are bath operators consisting of annihilation operators.
Under the rotating wave approximation, Eq. (1) is equivalent to the
Hamiltonian (1) in Ref. [2] with the noise parameters $\lambda ^{\left(
1\right) }=Re\left( g_l\right) ,\lambda ^{\left( 2\right) }=Im\left(
g_l\right) $, and $\lambda ^{\left( 3\right) }=0.$ So the Hamiltonian there
describes amplitude damping as well as phase damping. Most practical
decoherence belongs to this class, though the Hamiltonian is not in the most
general form. The coupling (1) in the comment can not be transformed into
the Hamiltonian describing pure dephasing without changes to the free
Hamiltonian of the qubits.

(ii) The FHE in Ref. [2] is designed with the assumption that we know
accurately the noise parameters $\lambda ^{\left( i\right) }$. These
parameters are determined by the type of the dissipation. If the noise
parameters do not vary with time, they can be detected by measuring suitable
observables of some test qubits subject to the same source of environmental
noise. To perform the detect, we may use a technique similar to quantum
tomograph of an unknown evolution of an open quantum system [4]. But FHE can
also be realized without the knowledge of the noise parameters. Here we
present a simple FHE technique having this advantage. We introduce a
classical homogeneous far-violet-detuned optical field which acts on all the
qubits. Under the adiabatic approximation, the Hamiltonian describing the
driving process has the form [5] 
\begin{equation}
\label{2}H_{drv}=-\stackunder{l}{\sum }\frac{2\left| g\right| ^2\left|
E\right| ^2}{\omega _{opt}-\omega _0}\sigma _l^z, 
\end{equation}
where $\omega _{opt}$ and $\omega _0$ are frequencies of the optical field
and of the qubits, respectively. By adjusting the intensity $\left| E\right|
^2$ of the optical field, we can choose the coefficient in Eq. (2) to
satisfy $\frac{2\left| g\right| ^2\left| E\right| ^2}{\omega _{opt}-\omega _0%
}=\omega _0$. The free Hamiltonian of the qubits is thus eliminated. This
FHE technique has another advantage, that is, it still works in the
circumstance $\lambda ^{\left( 3\right) }=0$, whereas the previous FHE
technique fails in this case.

(iii) As was shown in [1], the parameter $n_l$ in Eq. (10) of the letter [2]
is in fact zero. However, this restriction has no influence on the
construction of the universal gate operation, since the restriction is
indeed satisfied there (see Eq. (16) in [2]). The operator (14) constructed
in [2] is an encoded universal gate operation, not an encoded CNOT. It
reduces to CNOT only when the parameters $\alpha ,\theta ,\phi $ in $%
V_{l_2l_2^{^{\prime }}}\left( \alpha ,\theta ,\phi \right) $ (Eq. (15) in
[2]) have a special value, i.e., $\alpha =\phi =0,$ and $\theta =\frac \pi 2$%
. In contrast, for a universal gate operation, these parameters should be
irrational multiples of $\pi $ and of each other (see Ref. [2] and
references therein).

We note after acceptance of the letter [2] that a related work [6] was done
independently by Zanardi and Rasetti. They use four qubits, which are
required to be decohered collectively, to encode one qubit. The scheme there
deals with general dissipation, i.e., it prevents amplitude damping and
phase damping at the same time. The FHE is not needed in [6], but the scheme
is less efficient, and moreover, the condition of collective decoherence for
four qubits is not easy to satisfy in practice [7].\\

{\bf Acknowledgment}

We thank Zanardi and Rasetti for an advance copy of their comment. This
project was supported by the National Nature Science Foundation of China.\\

\end{document}